# Superferric CIC Dipole for JLEIC Design Report

DOE grant DE-SC0016243
Accelerator Research Laboratory
Texas A&M University
Prof. Peter McIntyre, PI

The Accelerator Research Lab (ARL) at Texas A&M University is developing a 3 T superferric cable-in-conduit (CIC) arc dipole for use in the Ion Ring of the proposed JLEIC. During the previous year ARL had prepared a full Conceptual Design Report [1] for the dipole and established that the magnetic, mechanical, and thermal design of the magnet should provide the performance requirements for JLEIC. Short-length samples of the CIC conductor were fabricated and tested for structural integrity, feasibility for bending to form the flared ends of the windings, and short-sample performance of the actual wires in the final form of the windings. A 1.2 m model of the dipole structure was fabricated, and a full winding model of the dipole was fabricated, using motorized bending tools to form the flared ends of the winding. The precise locations of the winding turns in the winding model were measured, and the random multipoles produced by cable position errors were calculated. The result of that study showed that the CIC dipole could be fabricated with the precision in conductor placement that would be required to produce a field homogeneity that met JLEIC requirements.

The technical objectives of the FY2016 project were:

- fabricate a long length of CIC cable, incorporating all features required for the CIC dipole.
- wind a few turns of the CIC cable onto the coil form (fabricated in FY15) and evaluate the coil-winding methods using CIC cable.
- Develop methods for splice joints and quench protection suitable for use in a 1.2 m model dipole and in 4 m JLEIC dipoles.

The tooling and fabrication procedures were developed to fabricate the 125 m cable lengths required for a 1.2 m model dipole. This required development of perforated center tube, a specialized cabling head for twist-pitch cabling of the 15-strand cable onto the center tube, and a drawbench suitable for the long draws of the sheath tube onto the cable. The following report details our FY2016 effort, and reports on several additional tasks that have been accomplished to bring the magnet development to readiness for fabrication and testing of a 1.2 m model dipole.

## CDC dipole design

Figure 1 shows the design for the 3 T CIC dipole, the error-field distribution at 3 T bore field, and a cutaway of the CIC cable. A key challenge for JLEIC requirements is to produce a dynamic aperture 10 cm x 6 cm (in which all multipoles have amplitudes $<10^{-4}$) over a range of field from 0.2 T to 3 T. The green-shaded regions in Figure 1b sarisfy that criterion.

The fabrication of short lengths of CIC cable is shown step-by-step in Figure 3. The 15-strand array of strands was wound with twist pitch by hand onto a center tube, it was wrapped with a thin stainless steel tape layer, it was inserted in a sheath tube, and the sheath tube was drawn down onto the cable to compress the strands against the center tube and immobilize them. The geometry was designed so that in the final drawn cross-section (Figure 3d) each strand contacts its neighbor strands and locally dimples slightly the center tube. The dimpled center tube provides immobilizing compression that is sustained through forming of the bends at the end windings and through cool-down (Figure 3e).

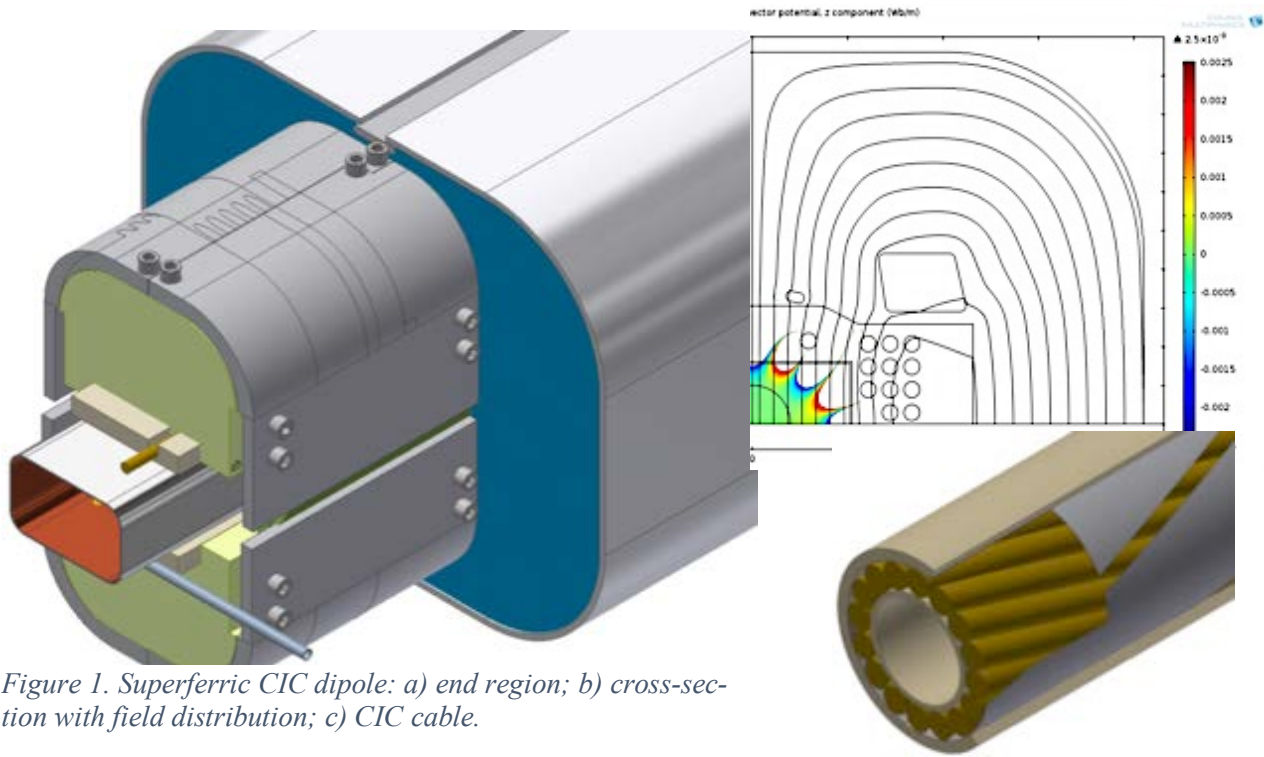

*Figure 1. Superferric CIC dipole: a) end region; b) cross-section with field distribution; c) CIC cable.*

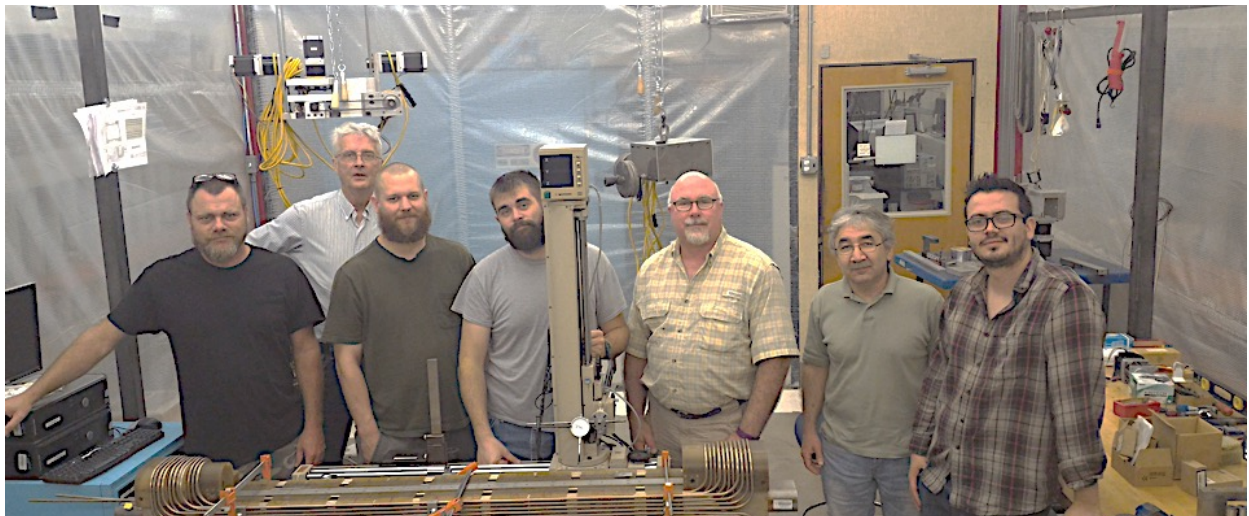

*Figure 2. Completed 1.2 m mockup winding model of the CIC dipole.*

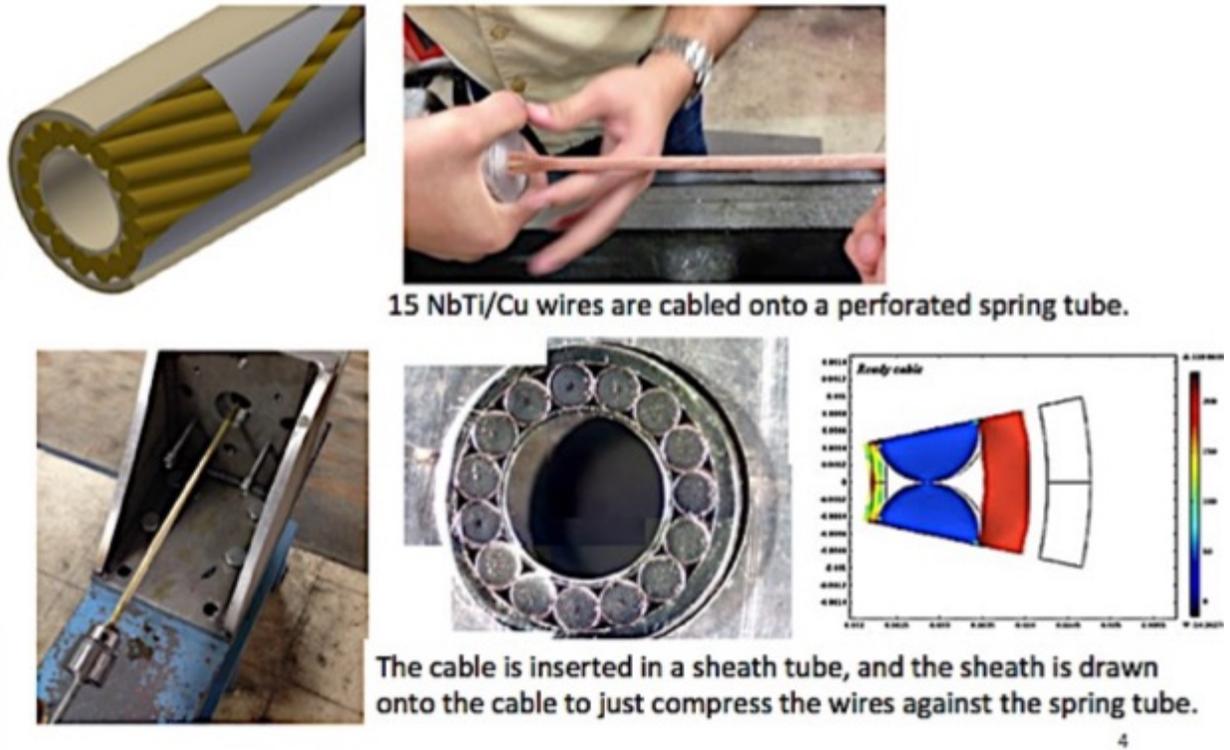

*Figure 3. Fabrication of short-lengths of CIC cable during FY15.*

## Development of long-length CIC cable fabrication
### Perforated center tube

The center tube must be made from an alloy that is non-magnetic, thin-wall (~.25 mm), and amenable to perforation. We devoted considerable effort and expense to develop a successful route for that purpose, ultimately custom-fabricated by a Chinese company [2]. Fabrication begins by slitting 0.25 mm-thick foil of 316L stainless steel to a width appropriate to roll-and-weld a tube of somewhat larger diameter than the desired 5 mm. The development of the process is shown in Figure 4. First the slit tape is perforated in a continuous reel-reel process (Figure 4a). Next the perforated tape is passed through a continuous tube-forming process in which is it formed into a tube and the seam is laser-welded (Figure 4b). The weld process worked intermittently at first – positioning the tape edges at the seam so that they were parallel and touching required precise tooling and adjustment. First attempts produced puckered regions at the weld. After improvement the process yields uniform welds (Figure 4c).

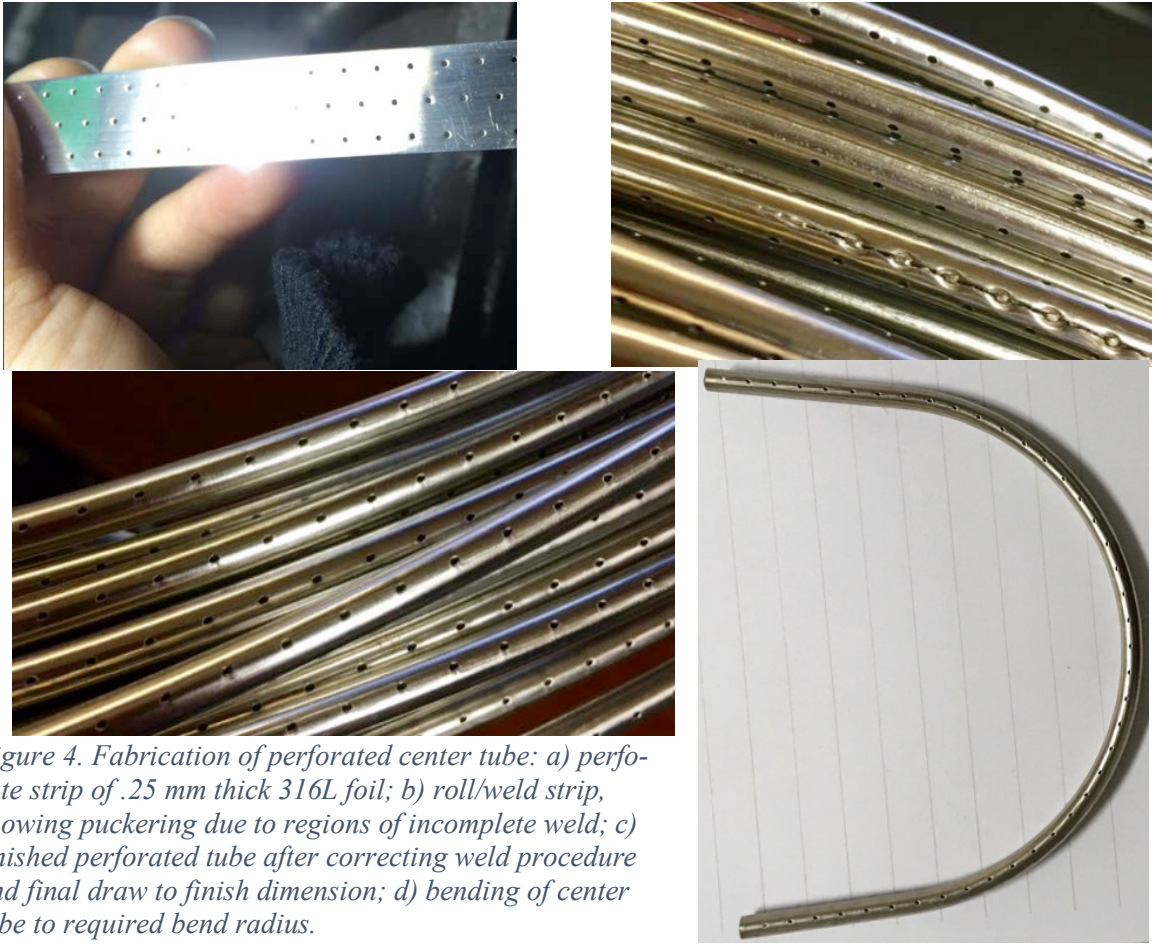

*Figure 4. Fabrication of perforated center tube: a) perforate strip of .25 mm thick 316L foil; b) roll/weld strip, showing puckering due to regions of incomplete weld; c) finished perforated tube after correcting weld procedure and final draw to finish dimension; d) bending of center tube to required bend radius.*

### Draw welded center tube to final OD to remove weld bulge

The center tube is then drawn down ~5% in diameter to remove the slight bulge in OD at the weld location. For this first requirement for long-length drawing we installed and commissioned a 12 m–long drawbench (Figure 5). We now have long-length quantities of perforated center tube.

We then formed short lengths of the welded tube into U-bends with the 5 cm radius required for the flared ends of the CIC windings. We measured the roundness of the center tube (Figure 4d), verified that we could limit ovaling to the desired degree, and verified that the welds were not compromised by the bending stress.

### Fabricate long-length CIC cable on perforated center tube

Long-length cabling requires continuous control of wire tension and twist pitch in the lay-up of the cabling operation. There are two options to accomplish that: contract cabling at New England Electric Wire, who have excellent machinery and experience in all forms of cabling; or development of a dedicated cabler for this particularly simple application. We opted for the latter approach for two reasons. First, we obtained budgetary quotes for both a single-run production of

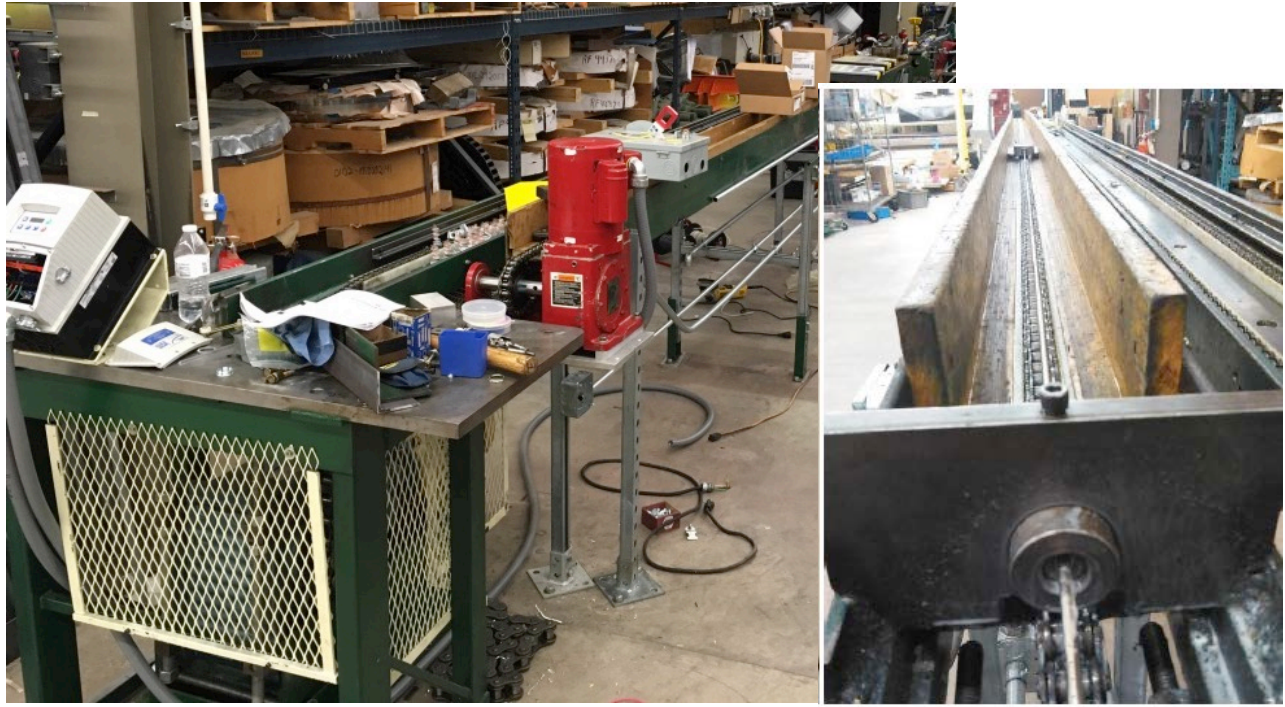

*Figure 5. a) 12 m long drawbench, used to draw center tube and sheath tube, and also to support the long-length cabling operation; b) drawing of perforated center tube to final OD.*

125 m cable for a 1.2 m model dipole, and for production of 500 such lengths for the full complement of 4 m arc dipoles that will be required for JLEIC. The single-run price was $4000 for one length; volume-run price was ~$3.85/m, corresponding to ~$1000/dipole.

Second, we became concerned that the requirement to spool the cable after cabling would then require a straightening operation as a pre-step for pulling the cable through the sheath tube. The spooling and straightening operations could produce risk of bunching of the stainless steel sheath wrap and de-registration of the cable.

We decided to instead develop a custom cabler, building upon what we learned in making short lengths. The long-length cabling operation is shown in Figure 6. The cabling head is shown in Figure 6d, mounted on the long-length drawbench. It is mounted on a rotary head that maintains constant twist pitch and constant tension as the head is traversed along the draw-bench. We have succeeded in fabricating 10 m cable lengths with spiral wrap of stainless steel overwrap (Figure 6c). The entire process is amenable to re-staging in the rear hall of our laboratory to extend the length to 125 m cable length.

### Fabricate long-length seamless tube for sheath

The CIC sheath tube provides the structural form for the cable, cable-level stress management during high-field operation in magnet windings, stable forming of flared ends, and containment of liquid helium for the cryogenic cooling of the windings. Those requirements have led us to evaluate several choices for the alloy of the sheath. Two candidates that look most attractive are Monel 400 and copper-nickel C70600. We are collaborating with HyperTech in development of a continuous tube rolling method (CTFF) that could directly form the sheath upon the cable and remove one process step in CIC manufacture (Figure 8). That development faces the challenge to show that CTFF tube can provide hermetic containment of helium at both room temp and at 4 K. Micro-cracks would produce leaks of LHe into the insulating vacuum. HyperTech has obtained encouraging success in Phase 1 development of CTFF cable in Monel 400, but we decided for the model dipole development to procure true-seamless tube in a size appropriate for drawing down on the cable.

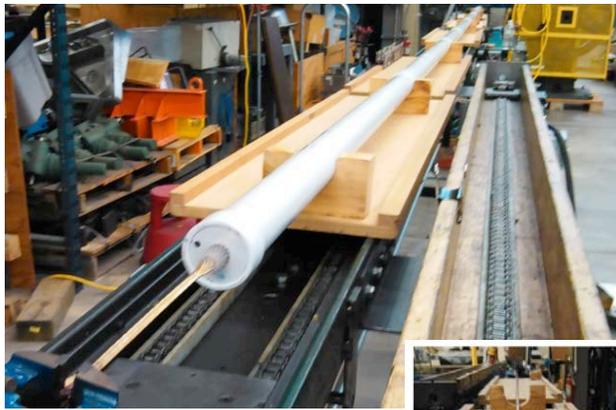
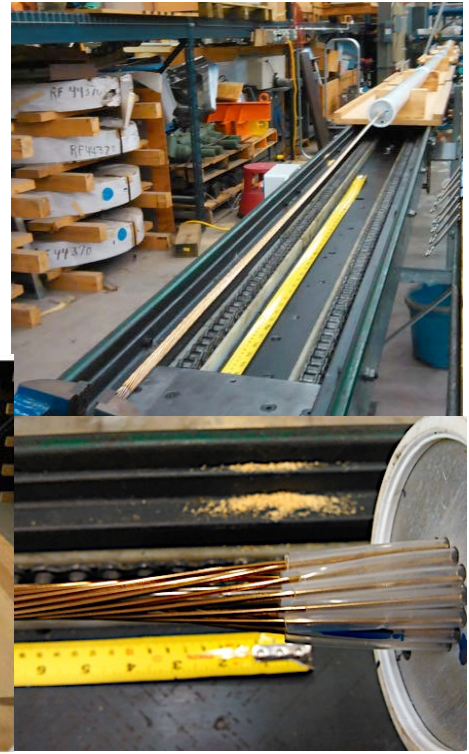
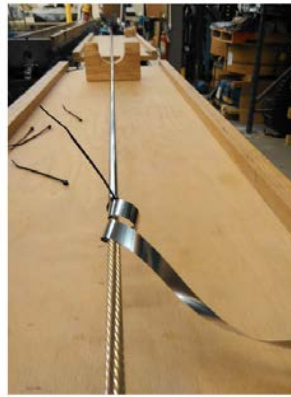

*Figure 6. Tooling for long-length cabling of 15 NbTi/Cu wires onto center tube: a) fixture with cabling die and sheathed bundle of long-length wire guides; b) tooling being transited along drawbench; c) over-wrap with SS tape; d) detail of cabling die.*

We procured true-seamless C70600 tube from Small Tube Products, and we ordered true-seamless Monel 400 tube from Shanghai Phoenix. The C70600 tube order was made successfully (Figure 7), and we have used it to make 10 m samples of CIC cable. The efforts to produce Monel 400 tube failed on first effort, and Shanghai Phoenix has recently succeeded with a second billet and tube-forming.

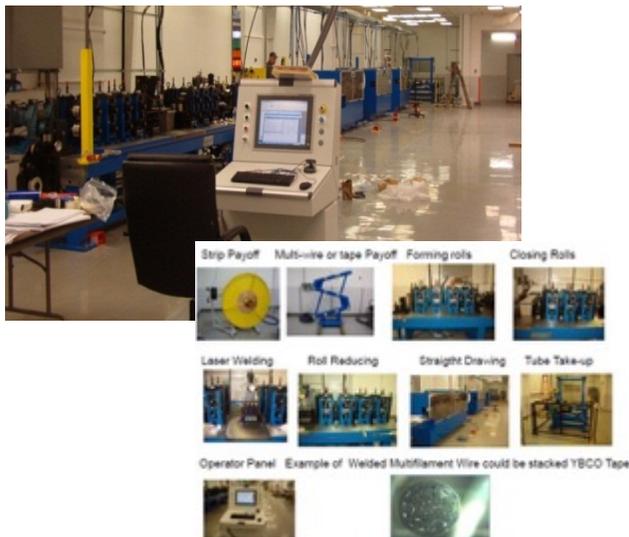
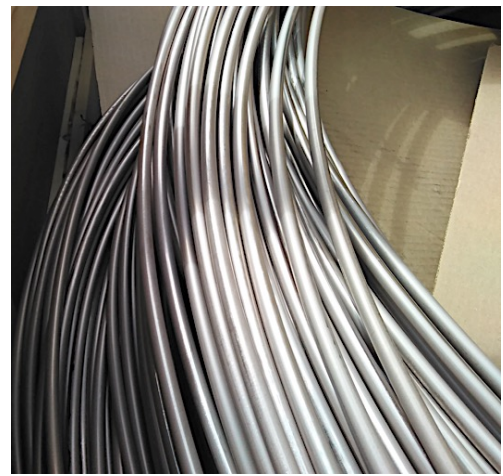

*Figure 8. Continuous tube forming of tube at HyperTech.*

*Figure 7. Seamless C70600 tube, 1 cm diameter, 0.5 mm wall thickness.*

## Success: Fabrication of long-length CIC conductor

Below is a photomontage of the sequence of operations to fabricate long-length CIC conductor.

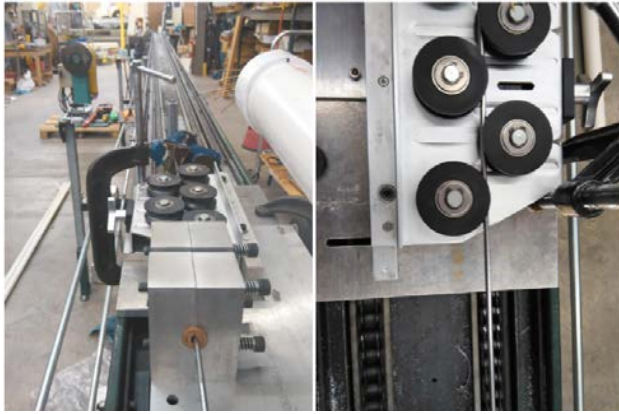
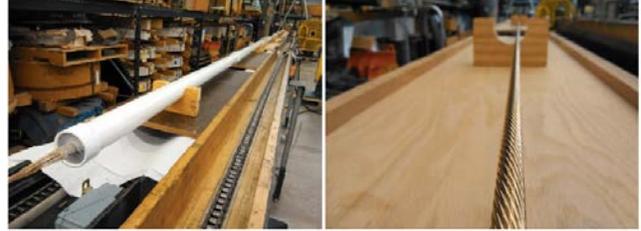
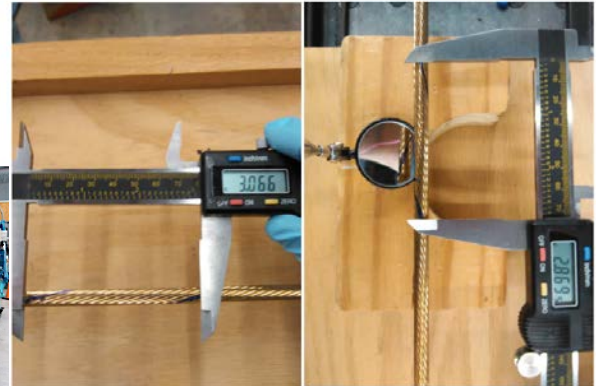
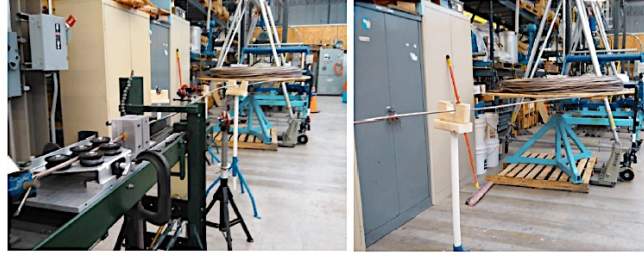
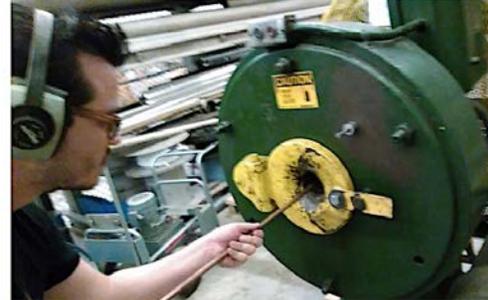
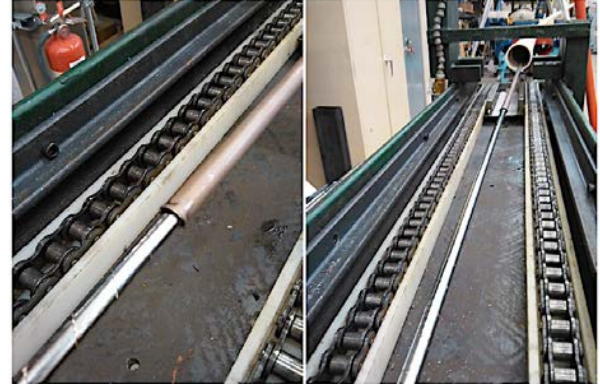
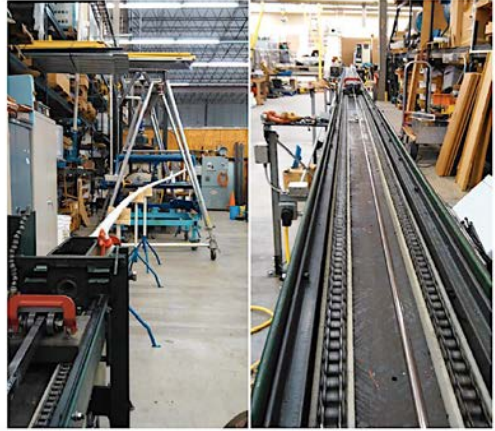
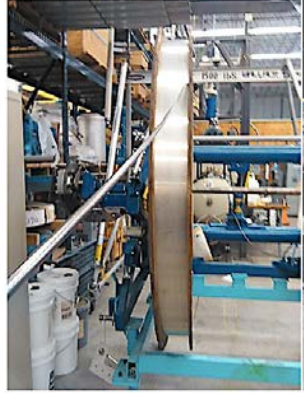

*Figure 9. Sequence of operations to fabricate long-length CIC cable*

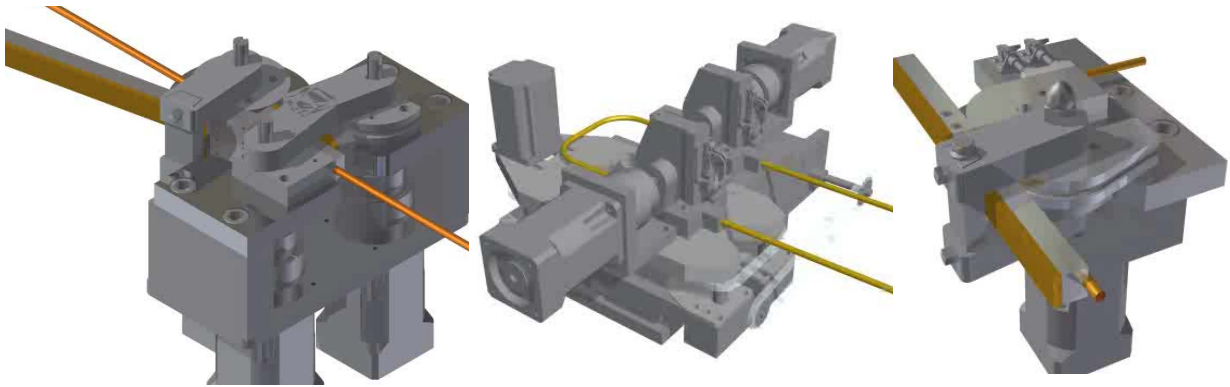

*Figure 10.Motorized bending tools developed to form the flared-end windings for the CIC dipole: a) bender to form 180° U-bend; bender to flare the U-bend 90°; c) bender to form 'dog-bone ends.*

### Form U-bends in CIC using motorized tooling

In our FY15 effort we developed motorized tooling to bend empty CuNi tube to the 5 cm radius required for the CIC end windings. We have adapted that tooling to similarly fabricate flared ends in actual CIC conductor.

The CIC conductor is much stiffer than the empty tube. We formed bends to determine whether the forming dies work correctly to bend CIC. We found that the bend requires more overbend to overcome spring-back. It will be necessary to fabricate a new set of bending dies must modify forming dies.

We were able to form U-bends to the required radius by shimming the present tooling, and we evaluated the bends by sectioning throughout the bend and extracting strands. The U-bends of CIC cable retained the cross-section and the strands remained intact and with full performance with no problems.

### Splice joint: how to connect CIC segments and manifold LHe flow

The 15 wires of each CIC conductor must be series-connected to a next CIC conductor (or to a flag of a current lead) with very low joint resistance (joint resistance of <nΩ is typical for Rutherford cable). Splice joints must be reliable and reproducible, they must have excellent heat transport to maintain them at the cold end of the operating temperature range for the cable. It should be possible to mount and de-mount a joint many times without degradation of the superconductor performance.

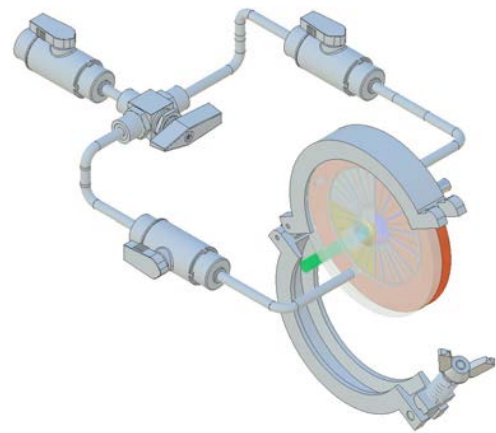

Splice joint technology is a significant challenge for any string of accelerator magnets, and can prove to be a major risk for overall performance if it is not reliable and robust –witness LHC.

*Figure 11. Design of splice joint suitable for interconnection of CIC conductors and parallel manifolding of LHe flow.*

We have developed a conceptual design for a splice joint that should provide such robust performance for CIC conductor in the JLEIC dipoles. It is shown in Figure 11. The main ingredient of the splice is a flare fitting that terminates each end of a CIC conductor. The flare fitting is made of OFHC copper, and its outer-exposed face is a disk with 15 radial channels milled into its end face. The channels are milled using a ball-end mill, so a NbTi/Cu wire can be compressed into the channel as a snug fit. The face of the fitting is pre-tinned: it is fluxed and then wipe-soldered with low-melt solder. The fitting also contains milled channels for LHe so that it flows out from the interior of the CIC into channels in the flare fitting, and then into a port on the flare fitting that will be used to manifold the flow between CIC segments and to provide inlet and outlet flows to the LHe supply manifolds.

Assembly of the joint proceeds as follows:
1. The flare fitting is slid over the sheath tube of a CIC conductor end, and the sheath tube is trimmed back to expose ~10 cm of the 15 NbTi/Cu wires. The sleeve joint between CIC sheath and fitting is soldered using low-melt solder.
2. Each wire end segment is flared to radial orientation; the 15 flared wire ends are oriented as a symmetric fan. The flared wire ends are pressed into their pre-tinned channels.
3. The fittings on two such CIC ends are pressed together with a thin gasket of fluxed low-melt solder between them. The two fittings are oriented so that the fans of NbTi/Cu wires are aligned with each wire of one CIC parallel and contacting a wire of the other CIC.
4. A clamp is installed on the pair of aligned flare fittings and tightened to compress them together. A cartridge heating jacket is energized on the clamp to rapidly heat the joint to melt and flow the low-melt solder and then cool it rapidly (to minimize bronzing of the Cu matrix in the wires).

We have simulated joint resistance in the CIC splice, and obtain ~0.1 n$\Omega$. We have simulated LHe flow through the joint and conclude that it should remain the coldest location in the dipole. In the FY17 continuation of our project we propose to build and test models of the splice joint.

# Topics that arose during FY16/17 requiring further study and analysis
## Quench heater design, simulation of quench dynamics

A 4 m 3 T MEIC dipole contains 24 turns of CIC cable, total cable length ~240 m. The dipole cross-section is shown in Figure 12. The CIC conductor is shown in Figure 1c. The cable is cooled by a flow of LHe through the center tube. The center tube is perforated so that LHe fills the void spaces among the strands. The dimensions of the strands and tubes are given in Table 1.

A resistive quench heater foil is located along a 10 cm segment at each end of the body length of the dipole, as shown in Figure 13. The serpentine heater foil is in thermal contact to each turn of CIC conductor. When a current pulse is passed through the heater foils quench is launched in both ends of every half-turn of the winding.

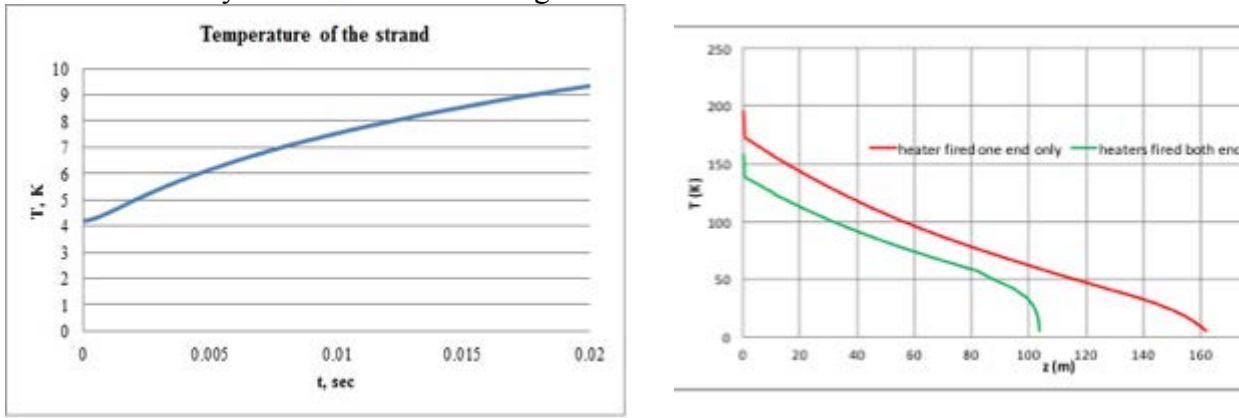

Figure 14. a) Time evolution of the quench, b) Temperature profile along each half-turn of CIC at end of quench.

We have simulated the propagation of a quench, at various values of cable current and locations of quench initiation. The quench begins at the quench heater and propagates along the cable with a velocity ~30 m/s. We have simulated quench propagation and heat deposition along the length of each half-turn cable length. Figure 14a shows the time development of a quench, and Figure 14b shows the temperature profile along a half-turn of cable. The case shown is for a worst-case scenario in which the quench heaters on one end of the magnet fail, so the quench is driven from one end.

Table 1. Dimensions of the CIC cable.

|  | OD | Thickness |
|---|---|---|
| Sheath tube | 8.25 mm | 0.5 mm |
| Center tube | 4.76 mm | 0.25 mm |
| Strands (15) | 1.20 mm |  |

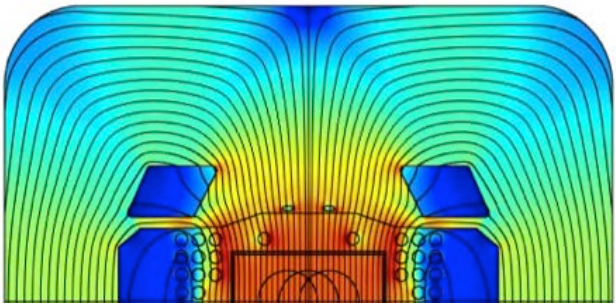

Figure 12. Cross-section of top half of MEIC dipole.

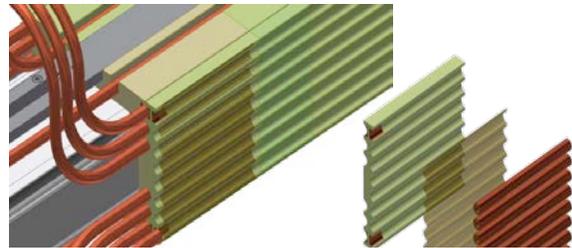

Figure 13. Resistive quench heater foil, located along a 10 cm length at opposite ends of the dipole.

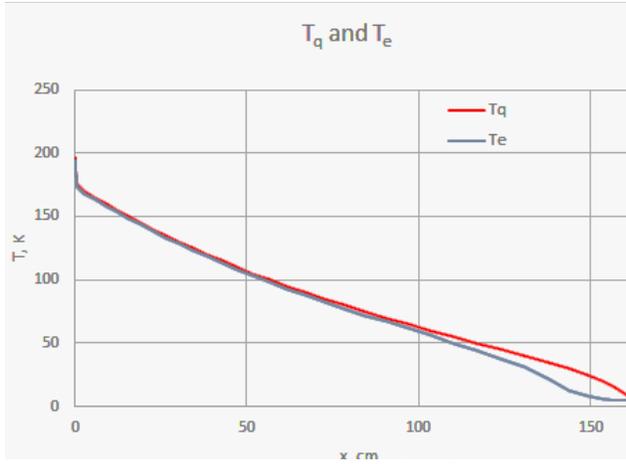
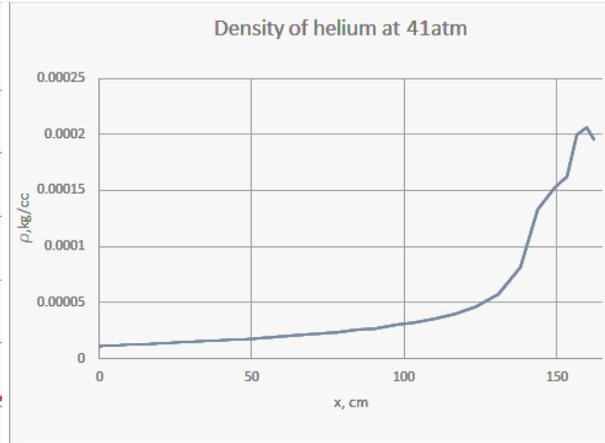

Figure 15. Cable temperature right after entire magnet quenches and at thermal equilibrium with helium.

Figure 16. Density of supercritical helium at local thermal equilibrium; pressure is 41 bar.

### Time constant for heat transfer from metal strands to LHe.

The ohmic heat in the quench is deposited in the metal of the CIC strands and tubes. The time constant by which heat transfers into the LHe within the CIC is $\tau = \frac{R^2 C_v}{Nu\,k}$, where $R = 2.5$ mm is the radius of the CIC center tube, $C_v \sim 0.8$ J/gK is the heat capacity of LHe, $k \sim 2 \times 10^{-4}$ W/cmK is the thermal conductivity of LHe, and $Nu$ is the Nusselt number associated with flow. The time constant is $\tau \sim 100$ s for static LHe in the CIC.

*Thus only a small portion of the heat transfers to the LHe during the ~0.2 s of a quench.*

### Flow velocity for re-distribution of LHe within the CIC during and following quench

Pending a cryogenics design for operation of the superconducting magnets in MEIC, we will assume that the liquid helium in the CIC conductors is in a supercritical state, with T ~4 K, P ~ 2 bar. Assuming a maximum local pressure of ~40 bar (we will return to this later), the flow velocity during quench will be

$$v_q = \sqrt{\frac{P_{max}}{L_{1/2}} \frac{2D_c}{f_D \rho}} \sim 70 m/s$$

Thus the helium locally within a 5 m half-turn of the winding will re-distribute internally to a change in the temperature distribution of the CIC conductor in a time that is short compared to the time in which the quench develops, so that *the pressure in the half-turn is homogeneous*.

### Evolution of pressure and temperature in the LHe

Quench is driven in all turns of the winding simultaneously, so the transients in pressure are ~periodic along the 48 half-turns of a magnet. In simulating the transient pressure response in a half-turn, therefore, it is a reasonable assumption that, during the development of a quench, no helium flows from one half-turn to the next: *the phase-temperature evolution of the LHe is an isochoric process.*

The CIC cable is supported within a modular G-11 matrix that is thermally insulating. Thus *the phase-temperature evolution of the LHe is an adiabatic process.*

center of the cable and assume that pressure surge due to hot helium will quickly displace low pressure, cold helium toward cold side of the cable equalizing the pressure in the cable to $p_0$.

The isochoric mass balance equation can be written as:

$$\int_0^{500cm} \rho_{He}(T_e(x), p_0)dx = [\rho_{He}(4.5K, 5atm.)][500cm]$$

The adiabatic local thermal balance can be written as

$$A_{cab}h_{cab}\left(T_q(x)\right) = A_{cab}h_{cab}(T_m(x)) + A_{He}\,\rho_{He}(T_{He}(x), p_0)\,[h_{He}(T_{He}(x), p_0) - h_{He}(4.5K, p_0)]$$

where $T_m(x)$ is the temperature of the metal right after the entire stored energy of the magnet is converted into thermal heat, $T_{He}(x)$ is the temperature which the metal and LHe of the CIC cable would approach (slowly) equilibrium when the LHe is at pressure $p_0$. $A_{cab}, h_{cab}$ are cable area and enthalpy of the cable, $A_{he}, \rho_{He}, h_{He}$ are the cooling channel cross section, helium density and its enthalpy. Tabulated specific volume *v* and enthalpy *h* of LHe were parameterized from NIST data as a function of temperature and pressure (summarized in the parameters of Table 2, Table 3, and Table 4. The enthalpy per unit volume (*h/v*) is a function only of pressure *p*, nearly independent of temperature.

The quench was simulated in the worst-case scenario, namely, quench heaters are fired only at one end of the dipole, and the slow transfer of heat from metal to LHe is assumed to happen quickly. As Figure 14b shows, the quench will dissipate ~235 kJ of heat (that half-turn's share of the stored magnetic field energy) in a length of ~3.2 m nearest the quench heater.

*The resulting LHe pressure is $p_0$ ~41 bar (4 MPa). Out of 235 kJ of thermal energy dissipated in the quench, 223 kJ goes to the metal and 12 kJ goes into the LHe process.*

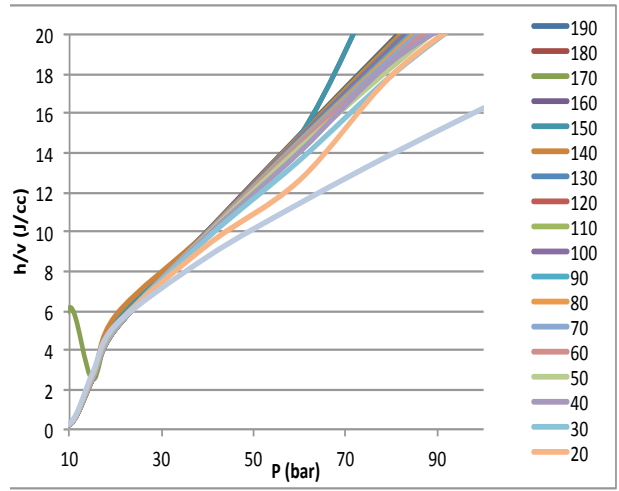

*Figure 17. Enthalpy/volume h/v as a function of pressure and temperature.*

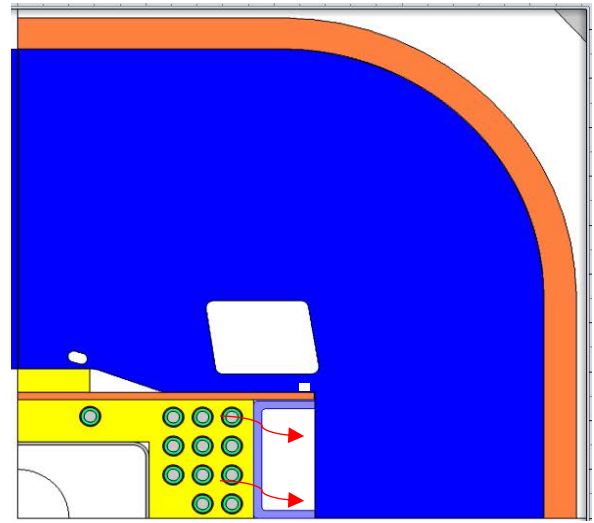

*Figure 18. Cross section of the dipole showing the Ti U-channel that would fill with LHe and He vapor in event of quench failure.*

Figure 15 shows the temperature distribution after quench along the half-turn in the metal ($T_q$) and in the LHe ($T_e$). Figure 16 shows the density of the LHe for the homogeneous pressure of 41 bar.

Both flow and heat transfer are rather complicated in supercritical helium [3]. We are still developing a simulation of the flow dynamics during quench in the periodically heated CIC winding. That dynamics may modify the above pseudo-isochoric estimations.

## Sheath tube stress from pressure surge

The pressure surge produces a hoop stress $S$ in the sheath tube:

$$S = P_{max} \frac{D}{t} = (4 \; MPa) \frac{8.25}{0.5} = 72 \; MPa$$

The yield strength at room temperature for the metals we contemplate as candidates for the sheath tube are:

|  | 300 K | 100 K |
|---|---|---|
| Copper-nickel 90/10: | 105 MPa | 150 MPa |
| Monel 400: | 170 MPa | 250 MPa |

There is thus ample strength for either choice of sheath tube alloy.

Note particularly that the above benign response is due to the intimate contact between the surfaces of all strands and the liquid helium. During a quench the pressure in the helium rises quickly above the triple point so that there is no vapor phase.

Even in that worst-case, the conductor would not be damaged.

## Ultimate failure mode: no quench protection

In the case that a quench occurred and the protection heater circuits failed to fire on *both* ends of the dipole, the local region where the quench began would heat to melt temperature and rupture its structure. Such a failure would destroy the winding, as is the case for *every* design of magnet for fields higher than ~1 T. The safety question in such a case is whether the LHe inventory of that dipole would then be contained within the internal space formed by the welded shell of the dipole.

A stainless steel compression shell (orange in Figure 17) is welded onto the outside of the dipole, to provide a stiff-wall closure of the gap between the two halves of the flux return under Lorentz loading. The shell hermetically seals the interior of the dipole so that helium ruptured from an un-protected quench would be confined within the cold mass enclosure. Ti U-channels flank the winding assembly; they are provided to control thermal expansion so that winding geometry is preserved through cool-down. Only the end regions of the winding assembly are vacuum-impregnated, so He from a rupture can flow through a pattern of holes into the interior void within the Ti U-channels, as shown in Figure 18. Liquid helium would remain liquid, and helium vapor that had boiled during the quench would be re-liquified in contact with the 4.2 K cold mass. Such a failure would present no safety issues.

*Table 2 Matrix of Coefficients for density parameterization*

| -3.2E-06 | 8.0E-04 | -5.9E-02 | 1.7E+00 | -2.2E+01 | 1.2E+02 | -2.4E+02 |
|---|---|---|---|---|---|---|
| 5.7E-07 | -1.0E-04 | 1.2E-02 | -3.5E-01 | 4.6E+00 | -2.5E+01 | 4.8E+01 |
| -3.7E-08 | 9.6E-06 | -7.5E-04 | 2.2E-02 | -2.8E-01 | 1.5E+00 | -2.8E+00 |
| 8.5E-10 | -2.2E-07 | 1.7E-05 | -4.8E-04 | 6.0E-03 | -3.2E-02 | 5.8E-02 |
| -5.9E-12 | 1.5E-09 | -1.1E-07 | 3.3E-06 | -4.0E-05 | 2.1E-04 | -3.8E-04 |

*Table 3 Matrix of coefficients for enthalpy parameterization*

| -4.4E+04 | 1.3E+04 | -4.3E+02 | 1.0E+01 | -1.3E-01 | 8.3E-04 | -2.8E-06 | 3.8E-09 |
|---|---|---|---|---|---|---|---|
| 6.1E+02 | -6.8E+02 | 4.8E+01 | -1.3E+00 | 1.6E-02 | -1.1E-04 | 3.9E-07 | -5.3E-10 |
| 5.9E+01 | 1.6E+01 | -1.4E+00 | 4.2E-02 | -5.7E-04 | 4.0E-06 | -1.4E-08 | 1.9E-11 |
| -1.5E+00 | -1.9E-01 | 2.0E-02 | -6.1E-04 | 8.5E-06 | -6.0E-08 | 2.1E-10 | -2.9E-13 |
| 1.0E-02 | 8.2E-04 | -1.0E-04 | 3.2E-06 | -4.5E-08 | 3.1E-10 | -1.1E-12 | 1.5E-15 |

*Table 4 Matrix of coefficients for cable enthalpy parameterization*

| 3.6E-02 | -4.2E-03 | 2.0E-04 | -1.2E-06 | 1.3E-09 | 6.8E-12 | 4.4E-15 | -7.1E-17 |
|---|---|---|---|---|---|---|---|

## 6 Tesla CIC dipole to double the ion energy in JLEIC?

The NP EIC Accelerator Technology Advisory Panel asked the question: 'What would be the feasibility and cost to double the energy of the Ion Ring of JLEIC using 6 T dipoles?'

We have endeavored to answer the question of feasibility for a CIC-based dipole. Figure 19 shows the cross-section of the dipole, and summarizes it's the main parameters of the 6 T design and the 3 T baseline design. As can be seen, doubling the dipole field can be accomplished with substantially the same CIC approach; it takes 2.25 x more conductor, and is somewhat larger. A more detailed design comparison would require a thorough design and prototyping of the high-field model. We can propose such development if it is of interest to JLEIC and/or DOE NP.

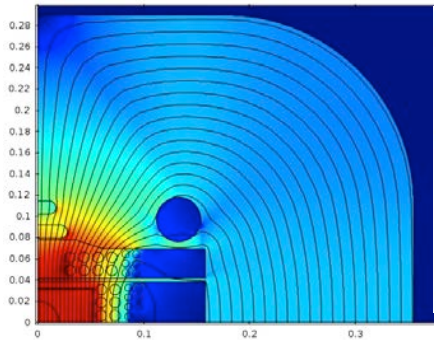

| Design field $B_0$ | 3 T | 6 T | 6T graded |
|---|---|---|---|
| Coil current | 13.7 kA | 17.2 kA | 18.6 |
| Coil field @ $B_0$ | 3.5 T | 6.9 T | 7.1 |
| Bore field @ SS | 3.8 T | 6.2 T | 6.4 |
| # turns in coil | 24 | 54 | 54 |
| Cable: | | | |
|   # strands | 15 | 14 | 18/10 |
|   strand dia. | 1.2 mm | 1.5 mm | 1.39 mm |
|   total s.c. area | 8 cm$^2$ | 27 cm$^2$ | 23 cm$^2$ |
| Flux return size | 20 cm | 33 cm | 35 cm |

**We significantly improved our earlier 6 T CIC design by grading the conductor.**

*Figure 19. CIC dipole designed to operate at 6 T.*

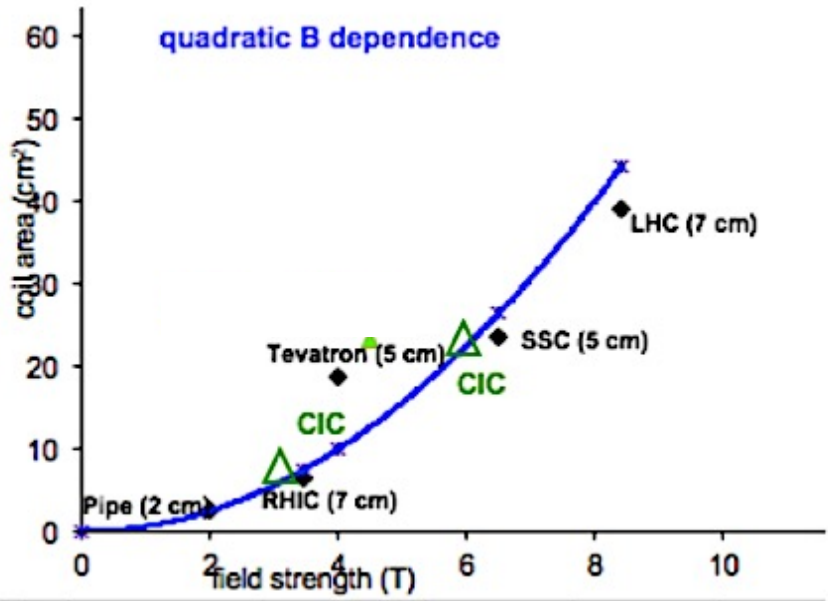

*Figure 20. Superconducting coil area vs. magnetic field for several winding geometries.*

### The CIC dipole is just as amp-efficient as cos θ coil dipoles

The windings of the CIC dipole are cables containing structure and LHe channels and the 15-strand NbTi/Cu cable. The fraction of cross-sectional area in the winding that contains superconducting wires is small. It has been suggested in various contexts that the CIC approach is less amp-efficient than cos θ dipole windings, in which a large fraction of the cross-section is devoted to superconducting wires.

Figure 20 presents the cross-sectional area of the windings in all of the superconducting dipoles ever used in accelerators and colliders, and also in several example designs for future colliders. The 3 T and 6 T CIC dipoles for JLEIC are shown as green triangles.

The superconductor required for a dipole depends ~quadratically upon the magnetic field in its bore. The only outlier from that dependence was the Tevatron, and that was because its superconducting wire predated the x2 revolution in current density.

***The superconductor area is ~independent of the coil geometry – cos θ, block-coil, pipe dipole, or CIC.*** It is simply not true that the structured coil of CIC cables requires more superconductor than a cos q dipole of the same bore field and aperture. It is just easier to build and more manufacturable.

### Flux plate to suppress persistent current multipoles and snap-back at injection

A collider ring must operate over a range of magnetic field from injection energy (when fresh beams are injected into its aperture) to collision energy (after acceleration to the energy for collider operation). When the magnet ring is ramped down after a collider store to injection energy, the discharge of the induction in the windings produces magnetization current loops within the superconducting filaments of every NbTi/Cu wire. The filaments are superconducting, and so the magnetization current loops are persistent. A pattern of macroscopic magnetic fields is produced by the superposition of the magnetization currents, and the induced field in the magnet aperture is rich in multipoles. This phenomenon of persistent-current multipoles is a significant problem for colliders. An even greater problem is the phenomenon of snap-back, in which the pattern of magnetization shifts suddenly when the current in the winding is changed from discharging to charging when the time comes to accelerate the new beam. The hysteresis of the magnetization response in the wires produces a sudden step change in the induction, and hence in the multipoles. This sudden change produces a sudden step in sextupole, hence a sudden step in chromaticity of the entire ring.

The amplitude of induced snap-back scales with the square of the diameter d of the filaments within each NbTi/Cu wire, and with the ramp rate $\dot{B}$ at which the dipole field must be increased on the charging curve. The Ion Ring of JLEIC requires a ramp time ~1 minute; the Booster (which will use shorter-length versions of the Ion Ring arc dipoles) requires a ramp time ~3 s. By contrast LHC requires a ramp time >300 s. Thus the persistent current fields will pose a significantly greater challenge for JLEIC than for previous superconducting colliders. While injection drift is slow enough to compensate, the snapback is difficult to predict or compensate.

We introduced a method of using a horizontal flat flux plate to suppress the persistent-current multipoles by placing two parallel flat strips of magnetic steel above and below the beam tube. At injection field ~0.2 T the steel is unsaturated and so the flux plates create a very strong dipole boundary condition that suppresses any field variations that are produced at the windings from propagating into the beam tube region.

An effective magnetization method [4] was be used to estimate maximum snap-back amplitude during the magnet ramp. Due to lack of magnetization measurements for JLEIC strands published data [5] were scaled and used in the simulations. To verify a proper scaling procedure, the experimental data of Figure 21 was compared at high field to the magnetization of the strand with fully penetrated by the magnetic field filaments

$$M(H,T) = \frac{2}{3\pi} d_{filament} \cdot \lambda \cdot J_c(H,T)$$

Where $\lambda$ isa fraction Non-Cu in the strand and $J_c(H,T)$ is a critical current density of the superconductor. A following parameterization [6,7,8] was used to estimate critical current density

$$J_c(B,T) = const \cdot B_{c2}(T)^{1.73} \left[\frac{B}{B_{c2}(T)}\right]^{0.948} \left(\frac{1}{B} - \frac{1}{B_{c2}(T)}\right)$$

$$B_{c2}(T) = 14.45 \left[1 - \left(\frac{T}{8.66[K]}\right)^{1.61}\right]$$

Figure 22 shows a parameterization of critical current density compared to the manufacturer specs of JLEIC strands. JLEIC strand magnetization (no transport current) was evaluated as

$$M_{JLEIC}(H,T) = \frac{\lambda_{JLEIC} \cdot d_{JLEIC}}{\lambda_{[1]} \cdot d_{[1]}} \cdot M_{[\text{Error! Reference source not found.}]}(H),$$

where $M, \lambda, d$ are magnetization, Non-Cu fraction and filament size (indices correspond to type of strands from Table 5), see Figure 23. Strands in the JLEIC cable were modeled as a $d_{strand}$ thick hollow cylindrical region and the magnetization was further scaled by a factor $\frac{A_s}{A}$, where A, $A_s$ are correspondingly total cross sectional area and strand cross sectional area (Figure 24).

Corrections to magnetization due to transport current density $J_{tr}$ is done using scaling factor $1 - \frac{J_{tr}}{J_c(H,T)}$. Finally magnetization of the hollow cylindrical region was calculated using following scaling

$$M(H,T) = \frac{\lambda_{JLEIC} \cdot d_{JLEIC}}{\lambda_{[1]} \cdot d_{[1]}} \cdot \frac{A_s}{A} \cdot (1 - \frac{J_{tr}}{J_c(H,T)}) \cdot M_{\text{[Error! Reference source not found.]}}(H)$$

The results of the modeling are given in Figure 25-Figure 27. Figure 25 gives multipoles due to strand magnetization, except of very low fields, but they are below injection fields ~0.2 T. Figure 26 shows multipoles when the magnets are continuously ramping down and up without injection stage. Figure 27 shows vector potential due to magnetization of the superconducting strands at 0.038 T and 0.19 T central fields. As one can see magnetization of the much needed at high fields sextupole correction coil creating substantial sextupole errors at low fields. To correct the errors 0.5 mm thick 85 mm wide steel flux plate was placed horizontally between the beam tube and correction coil (Figure 21) Position/size of the holes and shape of the iron yoke was re-optimized to correct field errors due to the flux plate (Figure 29) Vector potential due to magnetization currents at 0.05 T central field is shown in Figure 30, demonstrating how the flux plate improves homogeneity of the field due to strand magnetization. Show the sextupole (solid curves) and octupole (dashed curves) components of the fields induced by strand magnetization (Figure 31), and total multipoles (Figure 32) during ramping up (black colored) and down (red colored) of the winding current.

***The flux plate reduces persistent-current multipoles and snapback by a factor 5 compared to the same conditions without the flux plate.***

*Table 5 Strand parameters of measured and MEIC cable*

| type | Ref. 5 | JLEIC, Supercon Inc. |
|---|---|---|
| $d_{strand}$, mm | 0.648 | 1.2 |
| Cu/non-Cu | 1.3 | 1.5 |
| $J_c$ @4.2K, A/mm$^2$ | 2200 @5T | 1460 @7T |
| $N_{filaments}$ | 527 | 7400 |
| $d_{filaments}$, um | 23.2 | 9* |

\* estimated

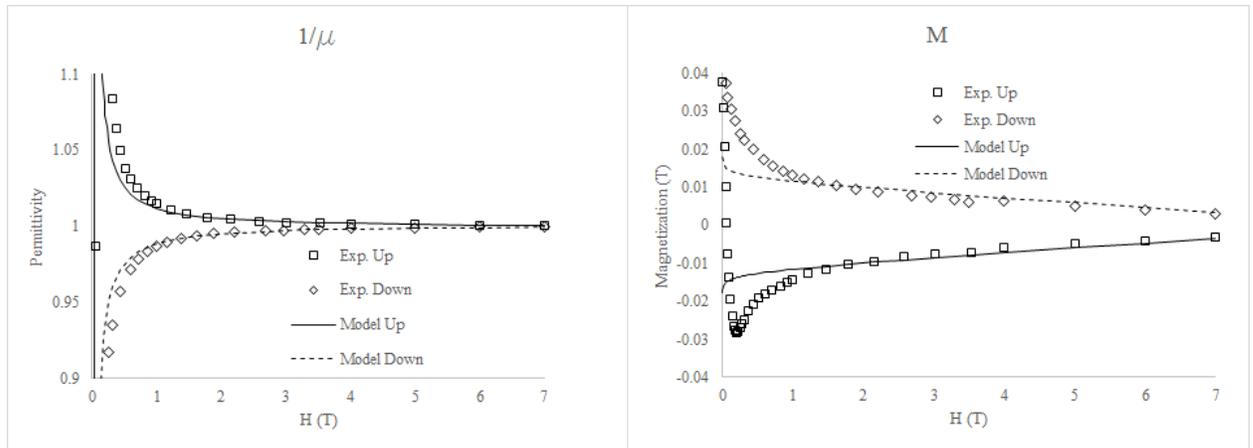

*Figure 21. Inverse of permeability and magnetization [5], solid and dashed lines are model results with magnetic fields fully penetrating all filaments of the strand.*

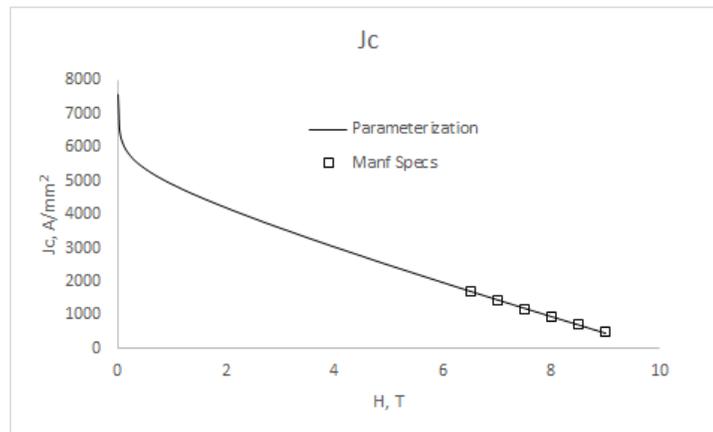

*Figure 22. Parameterization of $J_c$ compared to manufacturer specified values for JLEIC strands*

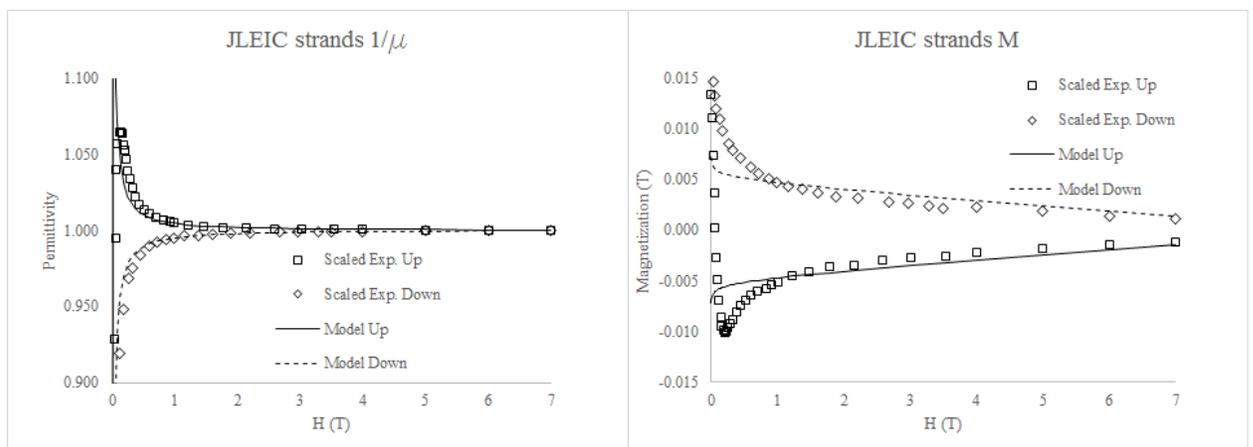

*Figure 23. Inverse of permeability and MJLEIC scaled from Ref. 5, solid and dashed lines are model results with magnetic fields fully penetrating all filaments of the strand*

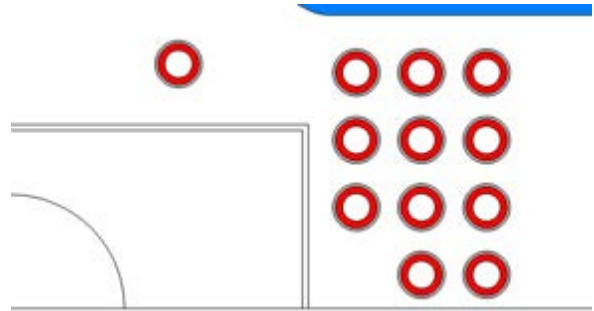

*Figure 24. JLEIC dipole cross section, strand occupied regions are shown in red*

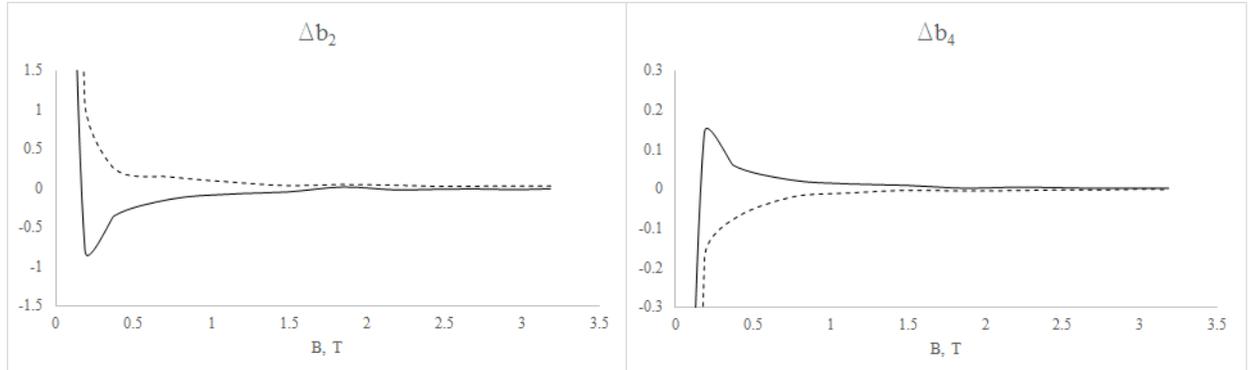

*Figure 25. Persistent current multipoles, solid line is while magnet is ramping up and dashed line is while the magnet ramping down*

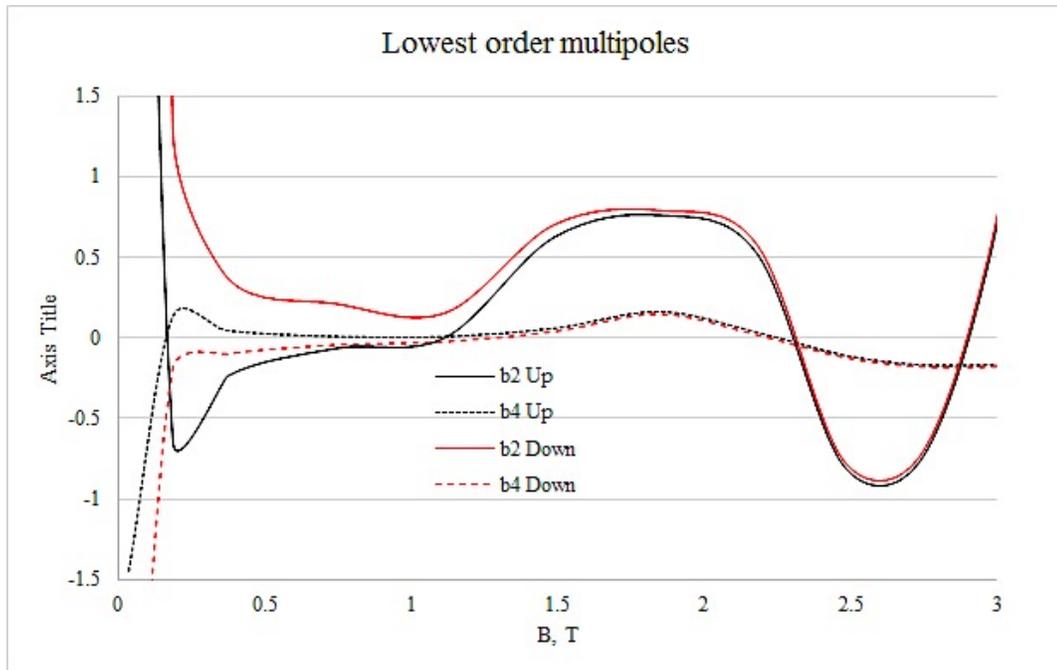

*Figure 26. Multipoles while ramping down (red curves) and ramping up (black curves)*

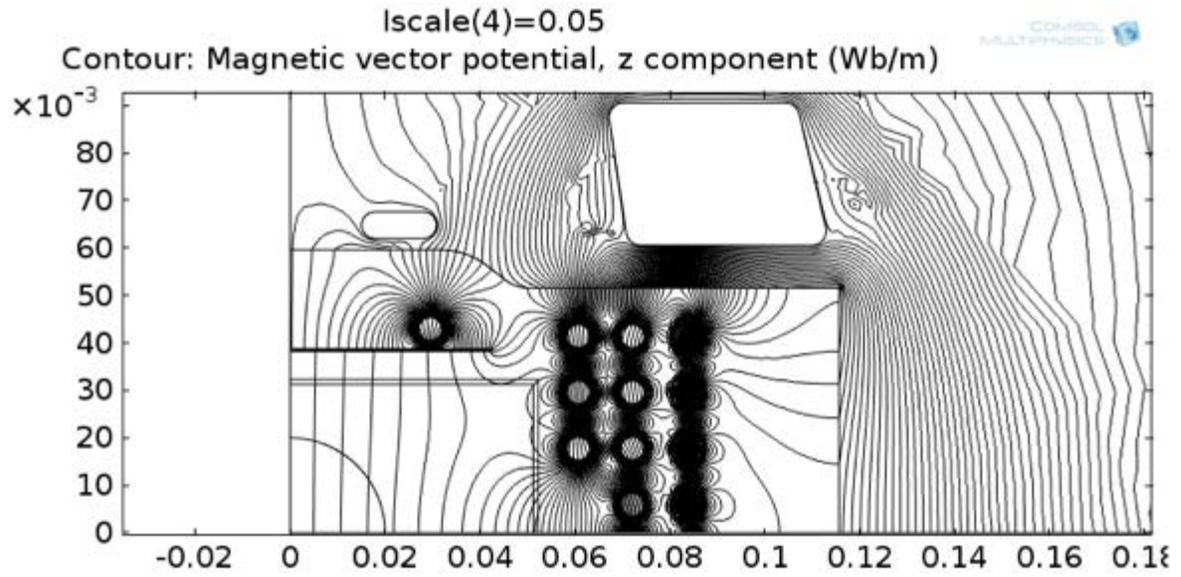

*Figure 27 Residual vector potential due to magnetization of superconducting strands at 0.05T.*

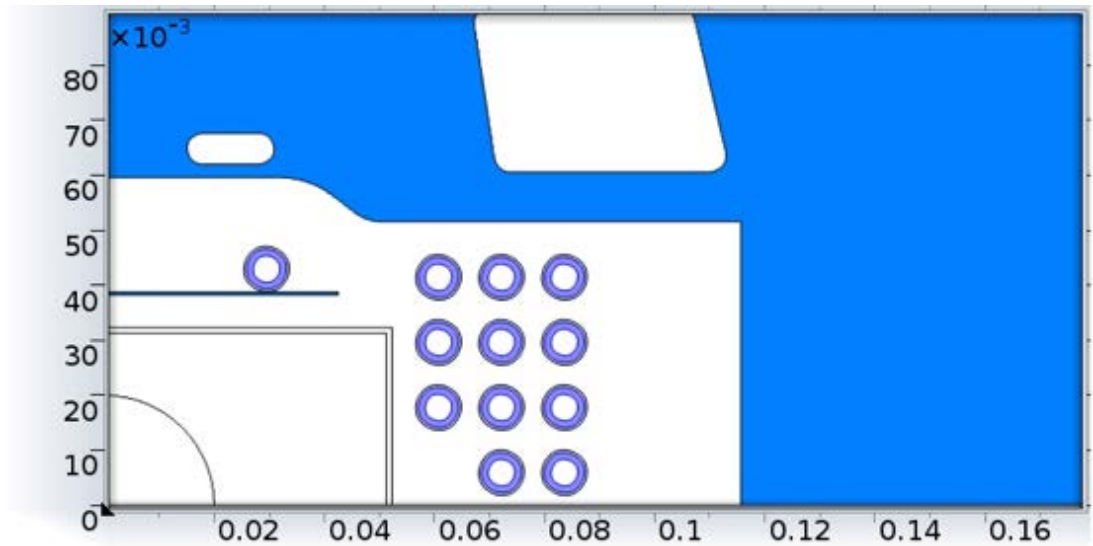

*Figure 28. 0.5mm thick 85mm wide horizontal flux plate placed between beam tube and sextupole correction coil*

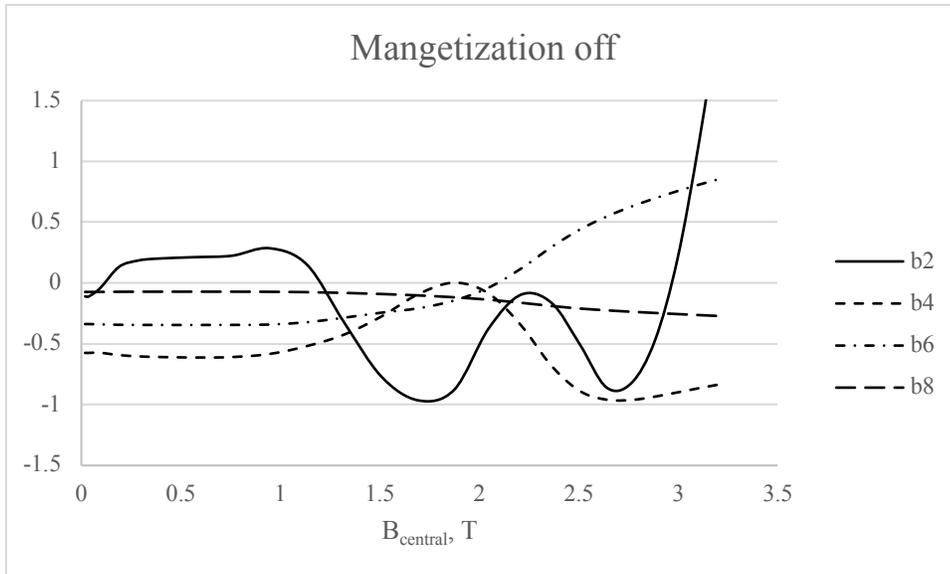

*Figure 29. Lowest order multipoles for re-optimized JLEIC dipole with persistent current induced multipoles suppressor*

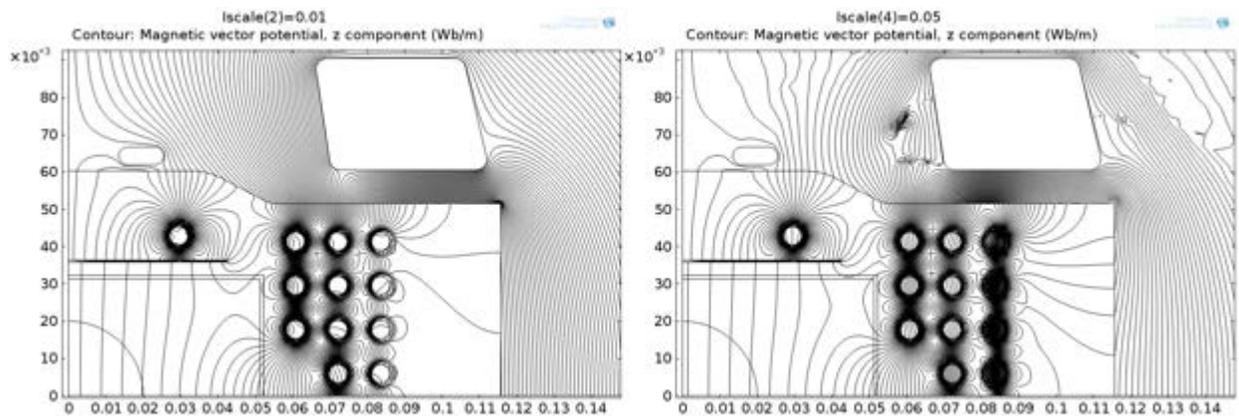

*Figure 30 Vector potential generated by magnetization currents at 0.04T (left) and 0.2T (right)*

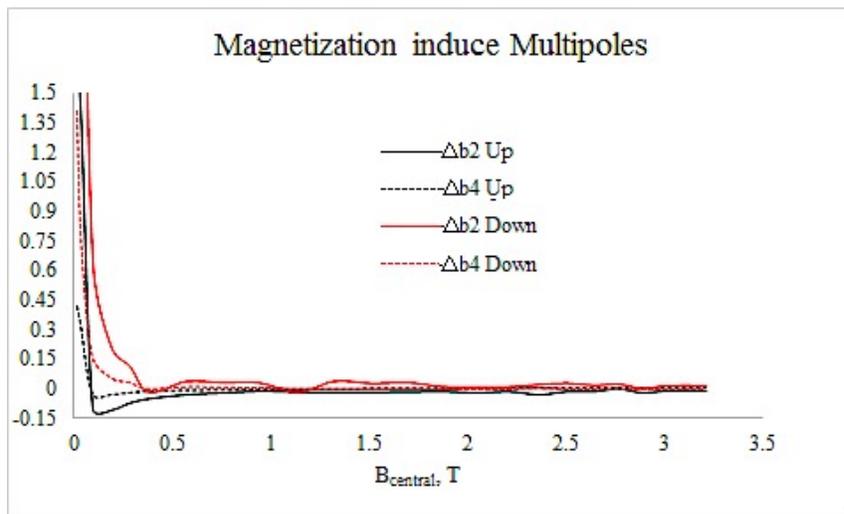

*Figure 31 Magnetization induced multipoles*

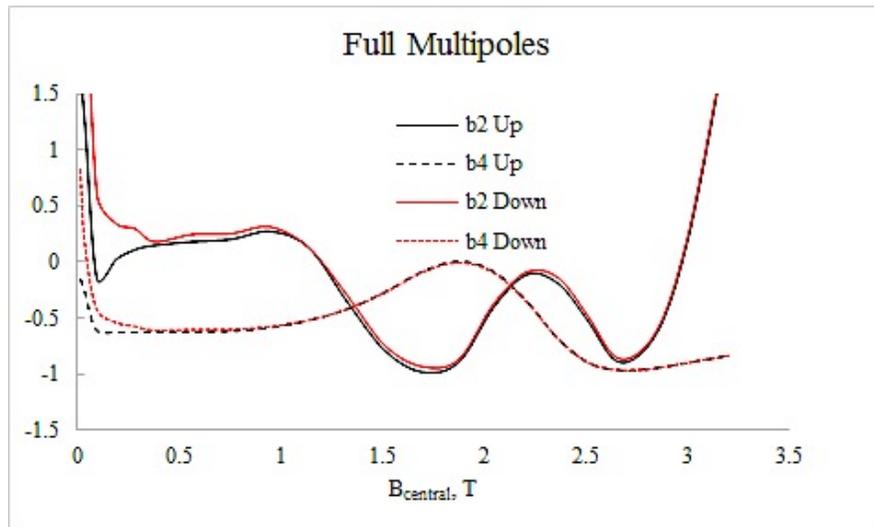

*Figure 32 Multipoles during ramping up and down of the JLEIC dipole.*

We have designed a modification of the FRP structural beam assembly that precisely positions all turns of the winding, to incorporate provisions for installing and aligning the flux plates. The modified design is shown in **Error! Reference source not found.**. It retains all of the structural rigidity of the original design and provides a precision-machined surface inside of the positions of the odd-man turns on which the flux plates are installed, the odd-man turns are installed, and positioning blocks are bolted to the structural beam to preload the turns and the flux plates against the structural beam.

## Summary

We have accomplished all of the technical objectives of the FY16-17 grant, and matured a basis of tooling and methods for cabling and drawing long lengths of CIC conductor. We have established that its properties are the same as those for the previous short-length CIC development.

We went on to address a succession of additional issues: quench dynamics is well-controlled under all failure modes; a 6 T 'doubler' is feasible with the same design methodology; and we can suppress persistent-current and snap-back fields by a factor 5 using a flux plate method that integrates naturally in the CIC geometry. This last feature is *only* possible with a block-coil geometry, and will likely be important in providing ramping performance required for JLEIC operation.